\documentclass{mn2e}
\usepackage{graphicx}

\def\msun{{\rm M_{\odot}}}

\title[The Short Period Supersoft Source in M31]
{The Short Period Supersoft Source in M31}
\author[A.R.~King, J.P.~Osborne, K. Schenker]{
A.R.~King, J.P.~Osborne, K. Schenker\\
Department of Physics and Astronomy, University of Leicester,
Leicester, LE1~7RH, UK}

\begin{document}

\maketitle

\begin{abstract}
We show that the recently discovered short period supersoft source in M31
is probably a progenitor of a magnetic CV. The white dwarf spins
asynchronously because of the current high accretion rate. However its
fieldstrength is typical of an AM~Herculis system, which is what it
will ultimately become. We discuss the relevance of this system to CV
evolution, and its relation to some particular CVs with special
characteristics.
\noindent

\end{abstract}

\section{Introduction}

A recent {\it XMM--Newton} observation of the central region of M31
(Osborne et al., 2001) has revealed a supersoft X--ray source XMMU
J004319.4+411758, hereafter M31PSS (= M31 periodic supersoft) with a
coherent period of $P = 865.5$~s. Periodicities in supersoft sources
have all hitherto been straightforwardly interpretable as
orbital. However the 865.5~s period of M31PSS is so much shorter than
the others ($P \ga 80$~min; Greiner, 2000) that this interpretation is
questionable. We examine in turn the ideas that the period is orbital,
or instead reflects the white dwarf spin.

\section{M31PSS as an ultrashort--period binary}

If the periodicity seen in M31PSS is orbital, the binary is
too tight to contain a main--sequence star, and must thus have a
degenerate companion. In this case the companion mass $M_2 = m_2\msun$
is given by
\begin{equation}
m_2 \simeq 1.5\times 10^{-2}(1+X)^{5/2}P_{\rm h}^{-1}
\label{m_2}
\end{equation}
and the mass transfer rate driven by gravitational radiation follows from
\begin{equation}
-\dot M_2 \simeq 1.3\times 10^{-3}(1+X)^{-20/3}m_2^{14/3}\ \msun {\rm yr}^{-1}
\label{mdot}
\end{equation}
(e.g. King, 1988). Here $X$ is the fractional hydrogen content by
mass, and $P_{\rm h}$ is the binary period in hours. A period of
865.5~s is below the minimum value for a hydrogen--rich secondary
(e.g. King, 1988), so the companion must be helium--rich ($X\simeq
0$). For M31PSS ($P_{\rm h} = 0.24$) we get $m_2 \simeq 0.0625, -\dot
M_2 \simeq 3\times 10^{-9}\ \msun {\rm yr}^{-1}$. There appear to be
four possible ways of attempting to explain M31PSS as an
ultrashort--period binary.

\subsection{M31PSS as a double--degenerate supersoft binary}

van Teeseling et al.\ (1997) suggested that RX J0439.8-6809 (whose
orbital period is unknown) might be a double--white dwarf supersoft
binary, on the basis of a very high X--ray to optical flux
ratio. Clearly this interpretation cannot work for M31PSS if it is a
member of M31, as accretion of helium--rich matter does not produce a
significant nuclear burning luminosity.

\subsection{M31PSS as a double--degenerate low--mass X--ray binary}

Since the nuclear luminosity of helium--rich accretion is too low to
power M31PSS if it is a member of M31, the only remaining way of
giving an ultrashort--period system the luminosity required for this
is to assume that the accretor is a neutron star or black
hole. 4U1820--30 (with a period of 11 min) is probably a system of
this type (Stella et al., 1987). However it is then difficult to
understand the observed supersoft spectrum. While this is natural for
a white dwarf accretor, where the characteristic effective temperature
$T_{\rm eff}$ is always close to $10^5$~K, neutron stars and black
holes have $T_{\rm eff}\sim 10^7$~K, giving emission peaking near 1
keV rather than 10 eV as in supersoft sources. In line with this, the
observed spectrum of 4U1820--30 is a 2.2~keV black body plus a power
law (Haberl et al., 1987), and thus completely unlike the supersoft
spectrum observed for M31PSS.

The next three possible explanations thus abandon the idea that M31PSS
is a member of M31, and assume instead that it is a foreground
Galactic source, and thus intrinsically much fainter.

\subsection{M31PSS as an AM CVn system}

The AM CVn systems are cataclysmic variables with He--rich secondary
stars and orbital periods of 17 -- 46 min, and so M31PSS could be a
member of this class if it is a foreground object.

The AM CVn systems are very faint X--ray sources, having luminosities
$L_{\rm x} < 10^{31}$ erg~s$^{-1}$ (van Teeseling \& Verbunt, 1994;
Ulla, 1995; Verbunt et al., 1997). However, they do not have supersoft X--ray
spectra. We have fitted archival ASCA data of the brightest AM CVn, GP
Com, and find that it is well described by optically thin emission at
$kT \sim 4$~keV, typical of non-magnetic CVs in general. Ulla (1995)
also does not find a supersoft X--ray spectrum from AM CVn itself.
The supersoft X--ray spectrum of M31PSS strongly distinguishes it from
the AM CVns.

\subsection{M31PSS as a double--degenerate AM~Her system}

Cropper et al.\ (1998) suggested that RX J1914.4+2456 might be a
double--degenerate AM~Herculis system. Interpreting M31PSS as a
foreground object in this way would differ from the AM~CVn idea above
in that accretion would occur along magnetic fieldlines to a
restricted region of the white dwarf. 

In this picture the white dwarf rotation is phase locked to the orbit,
so the modulation arises from occultation of the small bright
accretion spot by the body of the white dwarf as it rotates. The X-ray
light curve of RX J1914.4+2456 is indeed consistent with this
explanation, the X--ray flux going to zero as the small spot rotates
over the white dwarf limb. However the soft X--ray light curve of
M31PSS (Osborne et al., 2001) does not support this idea. Its simple
sinusoidal shape, with non--vanishing flux at all phases, is quite
unlike that of any of the known AM~Her systems, with the conceivable
exception of RX~J0453.4-4213 (Burwitz et al., 1996) which was observed
at fairly low signal--to--noise. Thus, to retain this explanation we
would require either that the accretion spot is unusually large,
and/or that we view the system from a very special orientation. Given
that we already require the source to be positioned quite by chance in
front of M31 it is clear that this is not a promising explanation.

\subsection{M31PSS as a foreground soft intermediate polar}

We conclude, in agreement with Osborne et al.\ (2001), that a more
plausible interpretation is that the period $P$ of M31PSS is the white
dwarf spin. This star must then possess a magnetic field strong enough
to channel the accretion flow, but too weak to lock the spin and
orbital rotations. There are two possibilities here: we examine below
the alternative possibility that M31PSS is intrinsically much
brighter, and actually a member of M31. If instead M31PSS has the
typical accretion--powered luminosity of known intermediate polars, it
cannot be a member of M31 but must again be a foreground object.

Intermediate polars with strong soft components are rare, but do exist
(Haberl \& Motch, 1995): currently 3 are known out of a total of more
than 20 intermediate polars. A problem for this interpretation is the
failure to detect an optical counterpart of M31PSS down to a limiting
magnitude of $B = 19$ in the December 2000 XMM--Newton Optical Monitor
observation in which M31PSS was faint in X--rays (Osborne et al.,
2001). Even assuming that the optical luminosity of M31PSS was
dominated by the companion star, with no accretion contribution, the
galactic latitude of M31 implies an implausibly large system distance
$z$ from the Galactic plane; we find $z \ga 350$~pc for an
intermediate polar with orbital period $\ga 5$~hr. Such a distance is
also required if M31PSS is not to be considerably fainter in soft
X--rays than the other intermediate polars with soft components. Given
that we are again requiring chance positioning in front of M31, it is
clear that this type of explanation is rather unlikely. 

\section{M31PSS as a supersoft intermediate polar in M31}

Given our conclusion above that the period $P$ of M31PSS is the white
dwarf spin, we are now left only with the possibility that the object
is a genuine member of M31, and is therefore intrinsically bright,
i.e. is indeed the supersoft source it appears to be. M31PSS is thus
the first known intermediate polar among supersoft X--ray
binaries. The latter systems consist of a white dwarf of mass $M_1$
accreting from a hydrogen--rich companion but differ from CVs in that
$M_2 \ga M_1$. This leads (van den Heuvel at al, 1992) to mass
transfer on the thermal timescale of the companion.

Thermal--timescale mass transfer rates are much higher than those in
CVs, which are usually driven by angular momentum loss. They lead to
the possibility of steady nuclear burning of the accreted matter on
the white dwarf surface. Assuming that M31PSS is a member of M31, the
X--ray luminosity implies that it has a typical supersoft X--ray
binary mass transfer rate $-\dot M_2 \sim 3 - 10 \times 10^{-8}\ \msun
{\rm yr}^{-1} \simeq 2\times 10^{18}\ {\rm g\ s}^{-1}$.

At the base of a steady white dwarf burning shell the pressure takes
values up to $10^{17} {\rm dyne}\ {\rm cm}^{-2}$ -- less than the
ignition pressure, but still much too large for magnetic confinement.
Once initiated, burning will rapidly propagate and consume any
hydrogen--rich material anywhere on the WD surface. However in a
supersoft system with a significant magnetic field, as we have
inferred for M31PSS, H--rich fuel steadily arrives at the magnetic
poles. Since the nuclear burning rate is strongly dependent on the
fractional H content, most burning will occur near the poles -- the H
content of the newly--arriving matter will drop severely as it flows
away from these sites, through the combined effects of burning and
geometrical dilution. The accretion rate needed to maintain steady
burning in this case is probably somewhat lower than in the
non--magnetic case, where the accreting matter is rather more spread
out in the disc boundary layer.

These considerations also allow us to rule out an alternative to the
steady nuclear burning identification -- a late stage of a nova
outburst.
Any asymmetry in the nuclear burning shell on a spinning WD could
provide the observed period of 865.5~s until the nova fades in
X--rays. However, our discussion of the previous paragraph shows that
in a nova, where no significant amount of new H--rich fuel arrives
after ignition, nuclear burning will quickly become spherically
symmetric; the slowly--accumulated fuel has had ample time to diffuse
over the entire white dwarf surface. Godon \& Shaviv (1995) show that
initial perturbations (e.g. local ignition at the poles) become global
on a dynamical timescale. Novae will therefore have spherically
symmetric nuclear burning and show {\em no} obvious azimuthal
inhomogeneity, and thus no periodicity. We conclude that the most
likely interpretation of M31PSS is as a supersoft intermediate polar.

\section{The Magnetic Field of M31PSS}

Thermal--timescale mass transfer of the type inferred for M31PSS is
possible without extreme assumptions only for orbital periods $P_{\rm
orb} \ga 5$~hr (King et al., 2001), and most observed periods are $\ga
10$~hr (Greiner, 2000).  This implies $P/P_{\rm orb} < 0.05$ for
M31PSS, suggesting that the system accretes from a fully developed
Keplerian disc rather than directly from an accretion stream, or a
magnetically influenced non--Keplerian disc (cf King \& Wynn, 1999).
In the Keplerian disc--fed case the specific angular momentum $j_{\rm
acc}$ accreted by the white dwarf is significantly less that of matter
leaving the companion through the inner Lagrange point ($j_{L1}$),
leading to an equilibrium with $P/P_{\rm orb}\la 0.04$, whereas one
gets $P/P_{\rm orb} \simeq 0.04$ in the stream--fed case with $j_{\rm
acc} = j_{L1}$ (King \& Lasota, 1991; King, 1993), and a whole series
of equilibria with $0.04 \la P/P_{\rm orb} \la 0.7$ in the
non--Keplerian case.

This identification of M31PSS constrains the magnetic moment $\mu =
BR_1^3$ of the white dwarf ($B=$ surface field, $R_1 =$
radius). Assuming this accretes from a Keplerian disc, rough equality
of magnetic and material stresses shows that the accretion flow will
be channelled by the field within a radius
\begin{equation}
R_{\rm M} \simeq 1.4\times 10^{10}\dot M_{18}^{-2/7}
m_1^{-1/7}\mu_{34}^{4/7}\  {\rm cm}
\label{rm}
\end{equation}
with $\dot M_{18} = -\dot M_2/(10^{18}\ {\rm g\ s}^{-1}), m_1 =
M_1/\msun$ and $\mu_{34} = \mu/(10^{34}{\rm Gcm^3})$ (see e.g. Frank
et al., 1992, eq. 6.10). The minimum requirement for channelling, i.e.
$R_{\rm M} > R_1 \sim 10^9$~cm, is satisfied for $\mu \ga 10^{32}{\rm
Gcm^3})$, or a surface field $\ga 10^5$~G.

A more stringent constraint comes from the assumption that the white
dwarf spin rate is close to the equilibrium value at which spinup via
accretion is balanced by centrifugal repulsion and other
torques. Magnetic accretors are likely to reach this equilibrium on a
timescale $\sim 10^5$~yr, i.e. much shorter than their evolution time
$M_2/(-\dot M_2) \sim 10^7$~yr. Variations in accretion rate on
timescales either $<<$ or $>> 10^5$~yr will still leave the spin close
to its current equilibrium value. The equilibrium assumption therefore
gives a reasonable estimate of the fieldstrength.

Assuming accretion from a Keplerian disc as before, equilibrium requires
$R_{\rm M}$ to be close to the disc radius where the local Kepler
period is the observed spin period $P = 865.5$~s, i.e.
\begin{equation}
R_{\rm M} = 1.5\times 10^8P^{2/3}m_1^{1/3} = 1.4\times
10^{10}m_1^{1/3}.\ {\rm cm}
\label{kep}
\end{equation}
Combining (\ref{rm}) and (\ref{kep}), with $\dot M_{18} = 2$, we see
that $\mu \simeq 1\times 10^{34}{\rm Gcm^3}$, or a surface field
$\simeq 1\times 10^7$~G. These values are typical of the AM Herculis
class of strongly magnetic CVs. The equilibrium condition (\ref{rm})
also implies a fractional accreting polecap area
\begin{equation}
f \simeq {R_1\over 2R_{\rm M}} \simeq 0.04
\label{f}
\end{equation}
(cf Frank et al., 1992, eq. 6.14). This relatively large value means
that the linear dimensions of the accretion region are a significant
fraction of the white dwarf radius, as is strongly suggested by the
approximately sinusoidal shape of the observed X--ray light curve.

Given an estimate of $\mu$ we can check the condition that the white
dwarf spin should not be locked to the binary orbit. As shown by
Hameury et al.\ (1987) this is equivalent to the requirement
\begin{equation}
2R_{\rm M} \la a,
\label{lock}
\end{equation}
where $a = 3.53\times 10^{10}P_{\rm orb,h}^{2/3}m_1^{1/3}$~cm is the
binary separation, with $P_{\rm orb,h}$ the binary period measured in
hours. This leads to
\begin{equation}
\dot M_{18} \ga 0.01\mu_{34}^2m_1^{-5/3}P_5^{-7/3}\ {\rm gs}^{-1}
\label{lock2}
\end{equation}
where $P_5$ is the orbital period measured in units of 5~hr. Since
we have $P_5 \ga 1$ (see above) this condition is easily satisfied. We
conclude that the identification of M31PSS as a supersoft intermediate
polar is self--consistent.

\section{The Future Evolution of M31PSS}

\begin{figure}
 \begin{center}
  \centerline{\includegraphics[clip,width=0.95\linewidth]{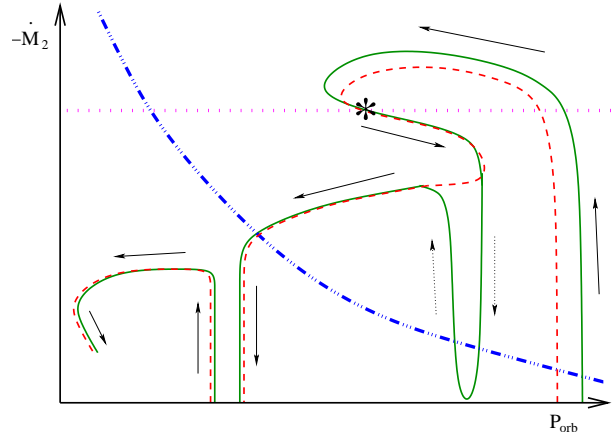}}
  \caption{Two possible scenarios for the future evolution of M31PSS,
  assuming its current state (marked *) is a magnetic
  thermal--timescale system with steady nuclear burning on the WD
  surface. Arrows indicate the direction of evolution; the dotted line
  is the current mass transfer rate in M31PSS, and the dash--dotted
  curve is the AM~Her condition given by (\protect\ref{lock2}).  The
  full curve shows a schematic evolution for very strong
  thermal--timescale mass transfer. The dip in $-\dot M_2$ after the
  period maximum marks the end of relaxation from the
  thermal--timescale phase. Relaxation began at the local period
  minimum, typically around 6--10~hr. For a slightly weaker
  thermal--timescale phase, the broken curve shows an evolution
  without such a drop. Both curves cross the AM Her line at short
  orbital periods (3--4~hr), but the first one also has a brief AM~Her
  phase around the period maximum (similar to V1309~Ori), typically
  8--12~hr. Both types of system spend the remainder of the evolution
  above the CV period gap as strongly magnetic, asynchronous WDs
  (i.e.\ possibly SW~Sex systems) rather than normal IPs.}
\label{fig:evol} 
\end{center}
\end{figure}

The thermal--timescale mass transfer driving the evolution of M31PSS
to its current state has increased the WD spin to the observed
value. Once the mass ratio has reduced so that $M_2 \la M_1$, mass
transfer will start to decline to a much lower rate driven by angular
momentum losses.  Provided that this latter value is $\ga 10^{-10}\
\msun {\rm yr}^{-1}$, (\ref{lock2}) shows that the system will become
a normal CV (note that $P_5 \ga 1$) with a strongly magnetic but
asynchronous white dwarf. Such systems are expected to be very
difficult to identify, as one can show (King \& Lasota, 1991, Section
IV) that both hard and soft X--ray emission will be strongly
suppressed. However the discovery of periodically varying circular
polarization in the SW Sex star LS Pegasi (Rodriguez-Gil et al., 2001)
may offer a clue as to their observational appearance. As the binary
period and mass transfer rate decrease further, the system will
eventually synchronize and become an AM~Her system (typically at about
3~hr, or below the period gap at 2~hr). Signs of its interesting
history would be a higher white dwarf mass than normal (the steady
nuclear burning during the supersoft phase makes $M_1$ grow by a few
tenths of a solar mass), and possibly some evidence of nuclear
processing in its abundances when the former CNO--burning core becomes
exposed. This evolution is shown as the broken curve in Fig.~1.

However for a sufficiently large initial mass ratio $M_2/M_1$ at the
start of mass transfer, the evolutionary tracks can show a pronounced
dip in mass transfer rate at the very end of the thermal phase (see
the full curve in Fig.~\ref{fig:evol}). The same phenomenon is found
by Podsiadlowski et al.\ (2001) in their study of LMXB evolution, and
is a result of relaxation back to thermal equilibrium.  As $-\dot M_2$
drops below $\sim 10^{-10}\ \msun {\rm yr}^{-1}$, the WD is able to
synchronize, and the system becomes an AM~Her at an unusually long
orbital period. The higher initial masses of the secondary star in
this scenario makes nuclear evolution possible, and the system may
show signs of CNO processing. A probable example of this case in our
Galaxy is V1309 Ori ($P_{\rm orb} = 8$~hr), which indeed appears to
have an overabundance of nitrogen (Schmidt \& Stockman, 2001)
indicating an originally much more massive CNO--burning main sequence
star. Once the companion is fully thermally relaxed the mass transfer
rate is likely to recover to more usual values ($\ga 10^{-9}\ \msun
{\rm yr}^{-1}$) for the binary period. The white dwarf will be spun up
and the system become asynchronous again. Eventually once the orbital
period has shortened to $\sim 3 - 2$~hr the white dwarf will
synchronize once more, leading to the second and final AM~Her stage of
the system's evolution.

In both the evolutions described here, the system may become a
propellor similar to AE~Aqr (Wynn et al., 1997) during spindown phases,
(e.g. when the white dwarf spin synchronizes) although the very short
spin period (33s) of AE~Aqr itself requires a weaker magnetic field
than in M31PSS. We note that AE~Aqr shows the strongest evidence in
any CV that nitrogen is enhanced at the expense of carbon (Jameson et
al., 1980). Curves like those in Fig.~\ref{fig:evol} can be matched to
the parameters of AE~Aqr (Schenker, 2001; Schenker et al., 2001).

In the December 2000 observation of M31, M31PSS had faded below
detectability. Moreover the lack of an optical identification means
that we have no information about masses or abundances. We are thus
currently unable to say which of the evolutionary paths sketched above
M31PSS will follow.  Either way it is clear that it is probably a
progenitor of a magnetic CV. Ironically we are largely prevented from
observing such progenitors in our own Galaxy because their short
lifetimes imply significant distances and thus heavy absorption;
conversely the distance to M31 means that CV descendants of systems
like M31PSS will be very hard to discover. This has the unfortunate
consequence that we cannot easily measure the relative numbers of CVs
and supersoft progenitors in an unbiased way. However both
considerations of the initial binary phase space, and the fact that we
see at least two probable supersoft descendants (AE Aqr, V1309 Ori) in
rather shortlived phases of their evolution as normal CVs, do suggest
that CV descent from supersoft sources must be relatively common.

\section{Acknowledgments}

Research in theoretical astrophysics and X--ray astronomy at Leicester
is supported by PPARC rolling grants. This research has made use of
data obtained from the Leicester Database and Archive Service at the
Department of Physics and Astronomy, Leicester University, UK.

\end{document}